\documentclass[preprint,aps]{revtex4}
\usepackage{graphics}
\usepackage{graphicx}
\usepackage{dcolumn}
\usepackage{bm}

\begin{document}
\title{Phonon-drag induced suppression of the Andreev hole current in superconducting niobium contacts}

\author{Kurt Gloos$^{1,2}$, Jouko Huupponen$^{1,2}$ and Elina Tuuli$^{1-3}$}

\address{$^1$ Wihuri Physical Laboratory, Department of Physics and
Astronomy, University of Turku, FIN-20014 Turku, Finland}

\address{$^2$ Turku University Centre for Materials and Surfaces 
  (MatSurf), FIN-20014 Turku, Finland}

\address{$^3$ Graduate School of Materials Research (GSMR), FIN-20500 Turku, Finland}



\begin{abstract}
We have investigated how the Andreev-reflection hole current at ballistic 
point contacts responds to a large bias voltage.
Its strong suppression could be explained by the drag excerted by the 
non-equilibrium phonon wind generated by high-energy electrons flowing
through the contact.
The hole - phonon interaction leads to scattering lengths of the low-energetic
holes down to 100\,nm, thereby destroying the coherent retracing of the 
electron path by the Andreev-reflected holes.
\end{abstract}

\maketitle 

\section{Introduction}
The Andreev reflection hole current at ballistic normal - superconducting point
contacts can be reduced by many mechanisms: normal reflection, finite Cooper
pair life time, and spin polarization produce characteristic anomalies and represent 
the material and device properties that we are  usually interested in \cite{Tuuli2011}. 
Others are associated with locally exceeding the superconducting 
critical temperature, field, or the condensation energy \cite{Gloos2009}. 
They result in sharp structures or side peaks of the spectra which can be easily 
reckognized.
However, they are difficult to handle because of their strong dependence on 
the geometric shape of the contact.
Another mechanism causes broad side peaks  at finite bias voltage 
that have neither well-defined  beginning nor end.
They add considerable uncertainty in the analysis of the spectra.
We suggest here that those broad maxima origin from the interaction of the 
Andreev reflection hole current with the wind of non-equilibrium phonons 
generated at large bias  voltages.

In a previous study on point contacts between ultra-high purity superconducting 
tantalum (Ta) and normal conducting silver (Ag), Hahn {\it et al.~}\cite{Hahn2002} 
have found  anomalies in the spectra that they interpreted in terms of a huge 
normal bubble - or 'hot spot' -  inside the superconductor. 
They assumed that high-energy electrons are resonantly trapped between the 
normal - superconductor interface of the hot spot itself and the contact.
Such a normal bubble would extend up to several times the superconducting 
coherence length of Ta. This length scale was supported by observing 
interference phenomena in a magnetic field \cite{Hahn1995}.
 
While Hahn {\it et al.~}\cite{Hahn2002,Hahn1995} were mainly interested in the 
resonances that appeared in the $dV/dI$ spectra to explain their observation, we 
have used a different approach based on the AR hole current derived from the 
difference of the measured normal and superconducting current-voltage
$I(V)$ characteristics of superconducting niobium (Nb) in contact with a normal 
metal.

\section{Experimental}

Our point contacts were fabricated using the shear method where the two sample
wires (typically $0.25\,$mm diameter) were brought into contact in the form of a cross. 
The differential resistance $dV/dI$ and the $I(V)$ characteristics were recorded in 
standard four-wire mode with low-frequency current modulation.

The thermal coupling between the contact, through the sample wires and their
anchorage, to the cooling stage at temperature $T_0$ was measured by injecting a 
large power into the contact and using the critical temperature $T_c = 9.2\,$K of Nb 
as marker for the contact turning normal.
This heating of the contact is described by $2R_{th} IV = L_0 \cdot (T^2 - T_0^2)$ with 
the Lorenz number $L_0$ and an equivalent thermal resistance $R_{th} < 1\,$m$\Omega$. 
An other upper bound for the local temperature distribution in the point contact 
is given by Kohlrausch's formula \cite{Kohlrausch1900} $V^2 = 4 L_0 (T^2 - T_0^2)$.
This is valid in the thermal limit, while in the diffusive and the ballistic regime the 
temperature differences are much smaller.

To measure the Andreev reflection excess current directly, we recorded the 
contact characteristics both  at low temperature (1\,K) in the superconducting and 
at high temperature (10\,K) in the normal state of Nb. 
This required thermal cycling until the contact stabilized and which worked 
well for normal-state contact resistances up to $R_N \approx 200\,\Omega$ or contact 
diameters down to around $2\,$nm.

\begin{figure}
  \includegraphics{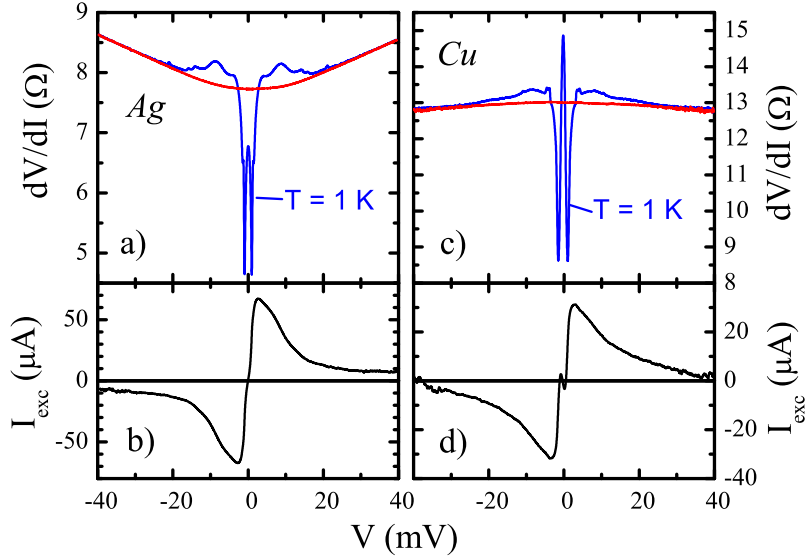}
  \caption{\label{spectra-1}
  $dV/dI$ versus $V$ spectra of a) a Nb - Ag and c) a Nb - Cu  contact at $1\,$K 
  and at $10\,$K. The corresponding excess current is shown in b) and d).}
  \end{figure}

\begin{figure}
  \includegraphics{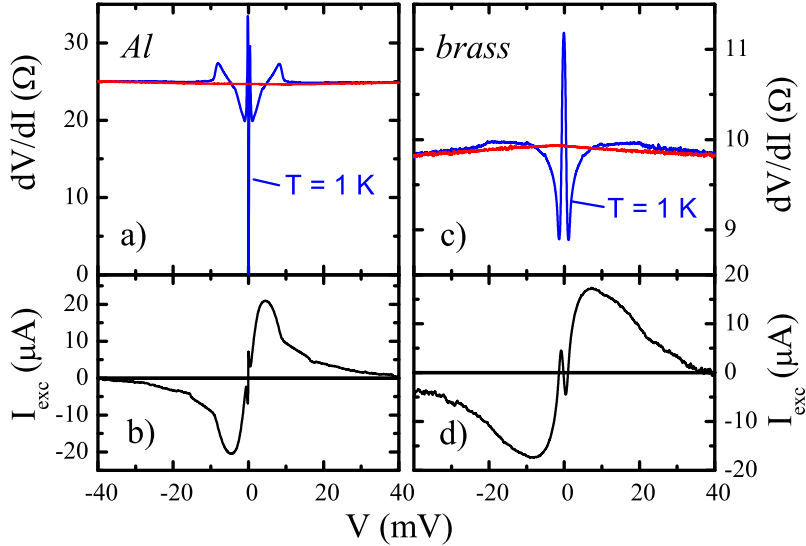}
  \caption{\label{spectra-2}
  $dV/dI$ versus $V$ spectra of a) a Nb - Al and c) a Nb - brass  contact at $1\,$K 
  and at $10\,$K. The corresponding excess current $I_{exc}$ is shown in b) and d).
  At $1\,$K Al is superconducting ($T_c = 1.2 K$) and the contact of Josephson type.
  However, the contribution of Al to the excess current is negligible compared to that
  of Nb.}
  \end{figure}

\section{Results}

Figures \ref{spectra-1} and \ref{spectra-2} show typical  $dV/dI$ spectra of 
Nb in contact with Ag, copper (Cu), aluminum (Al), and brass at $1\,$K and 
at $10\,$K as reference.
They look qualitatively similar to those observed by 
Hahn {\it et al.~}\cite{Hahn2002} for  Ta - Ag contacts. 
The spectra have the typical Andreev-reflection double-minimum anomaly
(except the contacts with Al) 
and also broad side peaks with a superposed fine structure.
At large voltages the spectra in the superconducting and the normal
state coincide.
The magnitude of the excess current $I_{exc}(V) = I_{1 K} (V) - I_{10 K} (V)$ 
drops continuously with increasing bias voltage which can roughly be described
by $I_{exc} (V) \sim exp{(-V/V_0)}$. 
Some of the contacts had a small residual excess current at large 
voltages, possibly due to a reversible change of the contact upon 
thermal cycling.
We have observed similar behaviour for all Nb - normal 
metal contacts. 
Figure \ref{V0} summarizes the decay constant $V_0$ of the excess current as 
function of normal-state contact resistance. 
It increases only slightly with $R_N$ and it apparently does not depend on 
the  normal metal.

\section{Discussion}

\begin{figure}
  \includegraphics{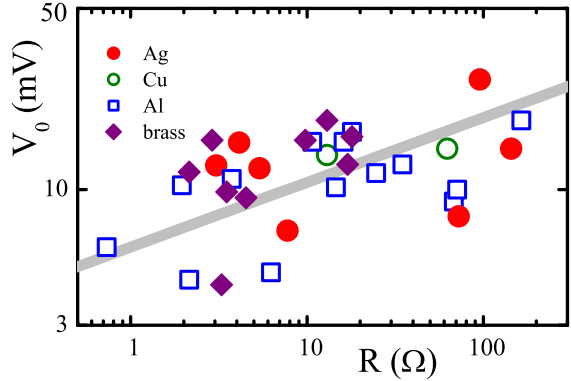}
  \caption{\label{V0}
    Decay constant $V_0$ of Nb in contact with the indicated metals as function of normal-state
    resistance $R_N$. The solid line is $V_0 \sim \sqrt[4]{R_N}$ as a guide-to-the-eye.}
  \end{figure}

At large bias voltage heating might become a problem. From the measured
thermal coupling of the samples we estimate maximum temperatures that are 
still far below $T_c$ of Nb.
The weak dependence of $V_0$ on the contact size also excludes heating effects,
because heating at a $R_N = 1\,\Omega$ contact would be 100 times stronger than 
at a $100\,\Omega$ contact, assuming the same bias voltage. 
This would contradict the experimental data in Figure \ref{V0}.

\begin{figure}
  \includegraphics{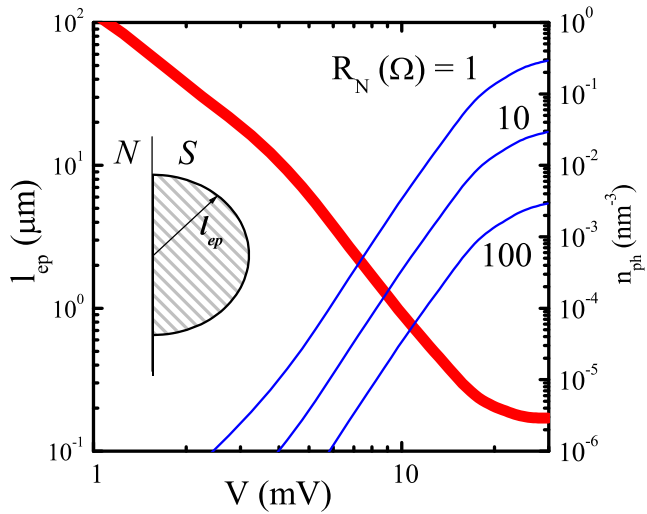}
  \caption{\label{EPI}
  Electron - phonon scattering length $l_{ep}$ as function of bias voltage $V$ 
  for Nb as derived from the point-contact electron-phonon interaction function
  and the excess phonon density $n_{ph}$ discussed in the text for contacts with 
  the indicated normal-state resistance $R_N$.
  Inset: Schematic of the normal (N) - superconducting (S) contact. The excess 
  phonon density is confined to a half-sphere of radius $l_{ep}$ around the 
  contact.}
  \end{figure}

To avoid the {\it ad hoc} assumptions of specific  boundary conditions used 
by Hahn {\it et al.~}\cite{Hahn2002}, we want to find a more conventional interpretation 
of the experiments. 
A rather straightforward mechanism to reduce the excess current without heating 
is electron scattering at the non-equilibrium phonons generated  by the 
bias voltage \cite{Jansen1980}.
Figure \ref{EPI} shows the calculated electron - phonon scattering length 
$l_{ep}(V) = v_F \cdot \tau_{ep}$ at low temperatures using 
$\hbar/\tau_{ep} = \int_0 ^{eV} \alpha^2 F(\epsilon)d\epsilon$ derived from the 
measured electron-phonon interaction function \cite{Khotkevich1995} 
and a Fermi velocity \cite{Ashcroft1976} $v_F = 1.37 \cdot 10^6$m/s of Nb.
At large voltages $l_{ep}$  is comparable to the superconducting coherence 
length $\xi \approx 50\,$nm of Nb. 
We estimate the density $n_{ph}$ of excess phonons as follows:
Phonons are generated at the point-like contact at a rate of $I/e$, 
that is one phonon per injected electron. 
The phonon density is constant inside a half-sphere of radius $l_{ep}$ 
around the contact.
They escape from this volume with sound velocity $v_s \approx 3500\,$m/s of 
niobium at a rate $2\pi l_{ep}^2 n_{ph} v_s$. 
Thus in the stationary state $n_{ph} = V/2\pi l_{ep}^2 v_s R_N$.
These excess phonons scatter the low-energy holes with a mean-free
path $l_{hp}$ comparable to that of the high-energy electrons $l_{ep}$
and destroy the coherence of Andreev reflection.
Such a scattering event deflects the hole from the path of the incident 
electron and thereby reduces its chances to fly back through the contact,
thereby reducing the Andreev-reflection excess current.
%

Since ballistic electrons have a preferred forward direction,
a cone would be more appropriate than a half-sphere. 
This would lead to a larger phonon density by reducing the volume and 
decreasing the area from which phonons can escape. 
In addition, each electron could create more than just a single phonon. 
Also the target area for the holes becomes  smaller with increasing contact 
resistance which could explain the very weak $V_0(R_N)$ dependence.

Our model also explains why some of the contacts have pronounced maxima and
side peaks while  others have very small anomalies. 
Large maxima of the type discussed here require the presence of an excess
current which can be suppressed, that is such a contact has little normal 
reflection and a small polarization as well as a long Cooper pair lifetime.
To extract those parameters from the measured spectra one should 
use the part around zero bias with a reasonable estimate of the normal contact
resistance, and ignore - as much as possible - the side peaks.


\section{Conclusions}

The suggested scenario offers no way to overcome the side-peak 
problem because generating non-equilibrium phonons 
is an  intrinsic property of the material and the contacts,
and inherent to measuring the spectra.
This makes a precise measurement of the energy gap of superconductors
with large $T_c$ and large $2\Delta$ difficult.
Side peaks can be avoided only when the superconductor has a small gap 
and the voltage range can be kept sufficiently small.

\section{acknowledgements}
We thank the Jenny and Antti Wihuri Foundation and the Magnus Ehrnroot 
Foundation for financial support.


\end{document}